\documentclass{iopart} 

\usepackage{graphicx}
\usepackage{epsfig,multicol,bbm}
\usepackage{amsbsy, setstack}

\newfont{\go}{ygoth.tfm scaled 1200}

\newcommand{\cwedge}[1]{\mathop{\wedge}_{{}^{#1}} }
\newcommand{\hook}{\raisebox{-0.35ex}{\makebox[0.6em][r]
{\scriptsize $-$}}\hspace{-0.15em}\raisebox{0.25ex}{\makebox[0.4em][l]{\tiny
 $|$}}}

\newcommand{\be}{\begin{equation}}
\newcommand{\ee}{\end{equation}}
\newcommand{\ba}{\begin{eqnarray}}
\newcommand{\ea}{\end{eqnarray}}

\newcommand{\eq}[1]{(\ref{#1})}

\newcommand{\mD}{{\cal D}}

\newcommand\fverb{\setbox\fverbbox=\hbox\bgroup\verb}
\newcommand\fverbdo{\egroup\medskip\noindent%
			\fbox{\unhbox\fverbbox}\ }
\newcommand\fverbit{\egroup\item[\fbox{\unhbox\fverbbox}]}
\newbox\fverbbox

\newcommand{\f}{f}
\newcommand{\ep}{\epsilon}

\begin{document}
\begin{flushright}
\small
DAMTP-2010-64 \\
\vspace{.3cm}
September 13, 2010 \\
\normalsize
\end{flushright}

\title{
Hidden symmetry in the presence of fluxes}

\author{David Kubiz\v n\'ak}
\address{DAMTP, University of Cambridge, Wilberforce Road, Cambridge CB3 0WA, UK}
\ead{D.Kubiznak@damtp.cam.ac.uk}

\author{Claude M. Warnick}
\address{DAMTP, University of Cambridge, Wilberforce Road, Cambridge CB3 0WA, UK
 and Queens' College, Cambridge CB3 9ET, UK}
\ead{C.M.Warnick@damtp.cam.ac.uk}

\author{Pavel Krtou\v{s}}

\address{Institute of Theoretical Physics, Faculty of Mathematics and Physics, Charles University, V Hole\v{s}ovi\v{c}k\'{a}ch 2, 180 00 Prague 8, Czech Republic}
\ead{krtous@mff.cuni.cz}

\begin{abstract}
We derive the most general first order symmetry operator for the Dirac equation coupled to arbitrary fluxes. Such an operator is given in terms of an inhomogenous form $\omega$ which is a solution to a coupled system of first order partial differential equations which we call the generalized conformal Killing--Yano system. Except trivial fluxes, solutions of this system are subject to additional constraints. We discuss various special cases of physical interest. In particular, we demonstrate that in the case of a Dirac operator coupled to the skew symmetric torsion and $U(1)$ field, the system of generalized conformal Killing--Yano equations decouples into the homogenous conformal Killing--Yano equations with torsion introduced in [arXiv:0905.0722] and the symmetry operator is essentially the one derived in [arXiv:1002.3616]. 
We also discuss the Dirac field coupled to a scalar potential and in the presence of 
5-form and 7-form fluxes.
\end{abstract}

\section{Introduction}

The Dirac operator, since its introduction in 1928, has played a central
role in physics and more recently geometry. Physically, the Dirac
operator is indispensable in the construction of the Standard Model of
particle physics. It is also intimately tied to Riemannian geometry;
indeed one approach treats the Dirac operator as fundamental and the
metric as a derived object \cite{Connes:book}. A fundamental tool of modern
physics is the identification and exploitation of symmetries. From a
metric point of view these are well studied, being related to the
existence of certain tensors, namely Killing tensors and Killing--Yano
tensors \cite{YanoBochner:1953}. Taking the view that the Dirac operator should take priority
over the metric, one may ask how the symmetries of the metric are
manifest in this alternative viewpoint. The key observation is that
the existence of symmetry operators for the Dirac equation implies
constraints on the geometrical background on which it is defined. In particular, symmetry
operators of the massless equation which are first order in derivatives are essentially in one-to-one
correspondence with conformal Killing--Yano tensors of the background
metric \cite{McLenaghanSpindel:1979, KamranMcLenaghan:1984, BennCharlton:1997, BennKress:2004}. 

In the backgrounds considered for superstring and supergravity
theories, the
metric is often supplemented by other fields or `fluxes' which couple
to the spinor fields and modify the Dirac equation. The goal of this
work is to understand how the symmetries of this modified Dirac
equation relate to the underlying geometry. This will give insight
into how the notion of `hidden symmetry' should be adapted in the
presence of fluxes. Generally, if $D$ is the standard Dirac
operator, one considers a modified Dirac equation:
\be
\mathcal{D} \psi = (D + B) \psi = 0\,, \label{moddir}
\ee
where $B$ is some section of the Clifford bundle\footnote{In the
  notation of physicists, $B$ may be written as a sum of terms of the
  form $B_{a_1 \ldots a_p} \gamma^{a_1}\ldots\gamma^{a_p}$, where $B_{a_1 \ldots a_p}$ are components of a $p$-form.}. This includes the case of a massive Dirac operator, the Dirac operator minimally
coupled to a Maxwell field, the Dirac operator in the presence of
torsion as well as more general operators. An interesting
question is under what circumstances this modified Dirac equation
admits a symmetry operator. A partial answer was provided by A\c{c}ik, Ertem, \"{O}nder and Ver\c{c}in
in \cite{AcikEtal:2009}, who give necessary and sufficient conditions that a first order operator {graded commutes} with
$\mathcal{D}$. If one asks only for a symmetry operator of
\eq{moddir}, it suffices to seek an operator ${\cal L}$ which {\em $R$-commutes} with
$\mathcal{D}$, i.e., which obeys 
\be\label{DLSD}
\mD{\cal L}=R\mD\,.
\ee
In this paper, we give necessary and sufficient conditions that a
first order differential operator on the spin bundle $R$-commutes with $\mD$.
As a result, we exhibit the
appropriate generalization of the conformal Killing--Yano equation in the presence of
fluxes, which we call the {\em generalized conformal Killing--Yano system}. This system incorporates all the special cases studied previously as well as providing a unified
description of some other possibilities.
In particular, we show how in the case where the fluxes consist of a
Maxwell field and a $3$-form torsion the conditions reduce to the
existence of a torsion conformal Killing--Yano tensor, as
introduced in \cite{KubiznakEtal:2009b}, together with a compatibility condition on the
Maxwell field. This reproduces in a more compact way the results derived in \cite{HouriEtal:2010a}  
and establishes the uniqueness of these results. As a new application we derive the symmetry operator for the Dirac field 
with arbitrary scalar potential and the generalization of 
a conformal Killing--Yano equation in the presence of 5-form and 7-form fluxes.

The paper is organized as follows.
In the next section we introduce the modified Dirac operator and discuss its basic properties. Its first-order symmetry operators are derived in Sec.~3. The general theory is demonstrated with several examples in Sec.~4.
Sec.~5 is devoted to conclusions and discussion. 
In what follows we use conventions and notations of \cite{HouriEtal:2010a} which we gather for convenience in the appendix.

\section{Dirac operator}
In what follows we consider the following Dirac operator:
\be\label{Dirac}
\mD=D+B=e^a\nabla_a+B\,,
\ee 
where $B$ is an arbitrary inhomogeneous form, describing the coupling to the `force fields'.
Since the standard Dirac operator $D$ is {\em antiself-adjoint}, i.e., obeying $D=-D^\dagger$, 
in order $\mD$ to be antiself-adjoint as well, $\mD=-\mD^\dagger$, we require  
\be
B=-B^\dagger=-(i)^{\pi(\pi-1)} B^*\,, 
\ee
where $*$ denotes the complex conjugation. This means that the $p$-form component of $B$, $B_p$, 
has to be imaginary for $p=0,1$ mod $4$ and real for $p = 2,3$ mod $4$.

An important property of the standard Dirac operator is that it obeys
the Schr\"odinger--Lichnerowicz formula relating its square to the
spinor Laplacian. We can derive a simlar result for $\mathcal{D}$, but
the spinor Laplacian related to $\mathcal{D}$ is that of a new
connection on the spin bundle. Let us introduce the following {\em spinor connection}:
\be\label{newconnection}
\tilde \nabla_a=\nabla_a+C_a\,,\quad C_a=\frac{1+\eta}{2} X_a\hook B+\frac{1-\eta}{2} e_a\wedge B\,,
\ee
where the action on a element of the Clifford algebra, $\alpha$, is
\be
\tilde \nabla_a \alpha = \nabla_a \alpha + C_a \alpha-\alpha C_a\, .
\ee
One can easily show that the Dirac operator \eq{Dirac} can be written as 
\be
\mD=e^a\tilde \nabla_a+\tilde B\,,\quad \tilde B=\Bigl[1-\frac{\pi}{2}(1-\eta)-\frac{n-\pi}{2}(1+\eta)\Bigr] B\,,
\ee
while we obtain  
\ba
\mD^2&=&\tilde \Delta +\frac{1}{2}e^{ab}\tilde{\cal  R}_{X_a,X_b}+(\mD \tilde B)\,. 
\ea
Here, $\tilde \Delta=\tilde \nabla_a\tilde \nabla^a-\tilde
\nabla_{\tilde \nabla_a X^a}$, a 2-form $\tilde{\cal  R}_{X_a,X_b}$ is
the curvature 2-form \eq{Rspinor} of connection $\tilde \nabla_a$, and in the last term $\mD$ acts only on $\tilde B$.
That is, by introducing $\tilde \nabla_a$, we have obtained the relation $\mD^2=\tilde \Delta+\mbox{zeroth order terms}\,.$
Such a connection is of interest on its own;
for example, let us consider the case when 
\be\label{AT}
B=iA-\frac{1}{4}T\,,
\ee where $A$ and $T$ are 1-form and 3-form, respectively. Then we find
\be
\mD=e^a\tilde \nabla_a+\frac{1}{2}T\,,\quad \tilde \nabla_a=\nabla_a+X_a\hook (iA-\frac{1}{4}T)\,.
\ee
This is a standard minimal coupling connection with torsion on spinors.
In the case when $A=0$, the connection $\tilde \nabla_a$ can in fact
be lifted from a connection $\nabla^T$ on the tangent bundle, as considered in \cite{HouriEtal:2010a}. On an arbitrary form $\omega$ this acts as
\be\label{covariantT}
\tilde \nabla_a\omega \equiv \nabla_a^T\omega =\nabla_a\omega+\frac{1}{2}(X_a\hook T)\cwedge{1}\omega\,.
\ee 
and one can associate with it the following two operations: 
\ba\label{def2}
\delta^T\omega &=&-X^a\hook\nabla^T_a\omega=\delta\omega-\frac{1}{2}T\cwedge{2}\omega\,,\nonumber\\
d^T\omega&=&e^a\wedge \nabla^T_a\omega=d\omega-T\cwedge{1}\omega\,. 
\ea
However, let us stress that this example is very special and in
general the spinorial connection $\tilde \nabla_a$, \eq{newconnection}, is not the lift of a connection on the tangent bundle because it
does not preserve the degree of forms.

\section{First-order Dirac symmetry operators}
We would like to construct the most general first-order differential operator ${\cal L}$
which graded R-commutes with the Dirac operator \eq{Dirac}, i.e., which obeys \eq{DLSD}. Equivalently,
we may define the {\em bracket} $\{\ ,\ \}$ to be
\be\label{bracketpaper}
\{\alpha,\beta\}\equiv\alpha\beta-(-1)^{q}\beta\alpha\,,
\ee
for a $p$-form $\beta$ and a $q$-form $\alpha$. Then one may show that $R$-commutation is equivalent to
\be\label{equation2}
\{\mathcal{D},{\cal L}\}= S \mathcal{D}\,
\ee
for some operator $S$.
Obviously, such an operator ${\cal L}$ is not unique. Given ${\cal L}$, a new operator ${\cal L}+\alpha \mD$, 
where $\alpha$ is an arbitrary inhomogeneous form, 
automatically satisfies the same equation \eq{equation2} (possibly with different $S$).
To construct all symmetry operators of $\mD$, it is in fact sufficient to seek a
special operator $L$, 
\be\label{L}
L=2\omega^a\nabla_a+\Omega\,,
\ee 
where $\omega^a$ and $\Omega$ are unknown inhomogenous forms to be determined,
obeying
\be\label{R}
\{\mD,L\}=\Sigma\mD\,,
\ee
where $\Sigma$ is of the zeroth-order, i.e., it is some inhomogeneous form \cite{BennKress:2004}.
A general operator ${\cal L}$ obeying \eq{equation2} is then given by ${\cal L}= L+\alpha \mD$, with $\alpha$ being an arbitrary inhomogeneous form.
In what follows we shall construct the special operator $L$.

Using the explicit form of $\mD$ and $L$, \eq{Dirac} and \eq{L}, we find 
\ba\label{DL}
\!\!\!\!\!\!\!\!\!\!\!\!\!\!\!\!\!\!
\{\mD,L\}=4X^a\hook \omega^b\nabla_{(a}\nabla_{b)}+2\Bigl(e^b(\nabla_b\omega^a)+X^a\hook \Omega+\{B,\omega^a\}\Bigr)\nabla_a
\nonumber\\
+2e^a\wedge \omega^b{\cal R}_{X_a,X_b}+e^a(\nabla_a\Omega)+\{B,\Omega\}-2(\eta\omega^a)(\nabla_aB)\,,
\ea
where we have used \eq{Rspinor}.
On the other hand, the r.h.s. of  Eq. \eq{R} is simply given by 
\be\label{RD2}
\Sigma \mD=\Sigma e^a\nabla_a+\Sigma B\,.
\ee
By equating terms of equal order in derivatives in Eqs. \eq{DL} and \eq{RD2} we obtain 
$X^{(a}\hook \omega^{b)}=0$, which implies
\be\label{omegaa}
\omega^a=X^a\hook \omega\,
\ee
for some inhomogeneous form $\omega$, together with 
\ba
e^b\nabla_b\omega^a+X_a\hook \Omega+\{B,\omega^a\}=\frac{1}{2}\Sigma e^a\,,\nonumber\\
2e^a\wedge \omega^bR_{X_a,X_b}+e^a(\nabla_a\Omega)+\{B,\Omega\}-2(\eta\omega^a)\nabla_aB=\Sigma B\,.
\ea
Using \eq{omegaa} and definitions \eq{dd} we have 
$e^b\nabla_b\omega_a=X_a\hook (\delta\omega-d\omega)+\nabla_a\omega\,.
$
Moreover, using the results of \cite{BennCharlton:1997} for an arbitrary inhomogeneous form $\omega$ one has 
\be
2e^a\wedge(X^b\hook\omega)R_{X_a,X_b}=e^b\wedge X^a\hook R(X_a,X_b)\omega\,.
\ee
So, besides \eq{omegaa}, we obtain the following two equations, which are the `conformal generalizations' of equations derived in \cite{AcikEtal:2009}:
\ba
\!\!\!\!\!\!\!\!\!\!\!\!\!\!\!\!\!\!\!\!\!\!\!\!\!\!\!\!\!\!\!
\nabla_a\omega-\frac{1}{2}e_a\wedge \eta \Sigma -X_a\hook\bigl(d\omega-\delta \omega-\Omega-\frac{1}{2}\eta \Sigma\bigr)+\{B,\omega_a\}=0\,,\label{eq1}\\
\!\!\!\!\!\!\!\!\!\!\!\!\!\!\!\!\!\!\!\!\!\!\!\!\!\!\!\!\!\!\!
e^b\wedge X^a\hook R(X_a,X_b)\omega+e^a(\nabla_a\Omega)+\{B,\Omega\}-2(\eta\omega^a)\nabla_aB-\Sigma B=0
\,.\label{eq2}
\ea
The first equation represents a ``{generalized conformal Killing--Yano system}'' of equations for inhomogeneous form $\omega$. It also uniquely determines the symmetry operator $L$. The second equation gives additional conditions which have to be satisfied in order that $L$ really be a symmetry operator for $\mD$. 

In order to see these statements more explicitly, let us first concentrate on Eq. \eq{eq1}. 
This can be re-written as 
\be\label{ckk}
\nabla_a\omega=X_a\hook\mu+e_a\wedge \nu -\{B,\omega_a\}\,,
\ee
where 
\be\label{ximu}
\mu=d\omega-\delta \omega-\Omega-\frac{1}{2}\eta \Sigma\,,\quad \nu=\frac{1}{2}\eta \Sigma\,.
\ee
By contracting \eq{ckk} with $X_a$ or wedging with $e^a$ we obtain the following two equations:
\be\label{ximu2}
d\omega=\pi\mu-e^a\wedge \{B,\omega_a\}\,,\quad 
-\delta\omega=(n-\pi)\nu-X^a\hook \{B,\omega_a\}\,.
\ee
By inverting these expressions we obtain 
\ba\label{ximu3}
\!\!\!\!\!\!\!\!\!\!\!\!\!\!\!\!\!\!\!\!\!\!\!\!\!\!\!\!\!\!\!\!\!\!\!
\mu=\frac{1}{\pi}d\omega+\frac{1}{\pi}\bigl(e^a\wedge\{B,\omega_a\}\bigr)-\f\,,\quad
\nu=\frac{1}{n-\pi}\bigl(X^a\hook\{B,\omega_a\})-\frac{1}{n-\pi}\delta\omega-\ep\,.
\ea 
Here, $\f$ and $\ep$ are arbitrary $0$-form and $n$-form respectively.
[We define $\pi^{-1}$ to be a linear operator taking $\alpha$ to
$p^{-1} \alpha$ for a $p$-form $\alpha$ with $p>0$ and similarly for $(n-\pi)
^{-1}$ acting on a $p$-form with $p<n$. Since $d\omega$ has no
$0$-form component and $\delta\omega$ no $n$-form component, the
expression above is well defined.] 
Using the explicit form of $\mu$ and $\nu$, \eq{ximu}, in \eq{ximu3} we obtain the following explicit expressions 
for $\Omega$ and $\Sigma$:
\ba
\!\!\!\!\!\!\!\!\!\!\!\!\!\!\!\!\!\!\!\!\!\!\!\!\!\!\!\!\!\!\!\!\!\!\!
\Omega =\frac{\pi-1}{\pi}d\omega-\frac{n-\pi-1}{n-\pi}\delta \omega-\frac{1}{n-\pi}X^a\hook \{B,\omega_a\}-\frac{1}{\pi}e^a\wedge\{B,\omega_a\}+\f+\ep\,,\quad\label{Omega}\\
\!\!\!\!\!\!\!\!\!\!\!\!\!\!\!\!\!\!\!\!\!\!\!\!\!\!\!\!\!\!\!\!\!\!\!
\Sigma=-\frac{2\eta}{n-\pi}\delta \omega+\frac{2\eta}{n-\pi}X^a\hook \{B,\omega_a\}-2\eta\ep\,.\label{sigma}
\ea
Moreover, let us introduce the following projection operator for inhomogeneous forms $\alpha_a$, see, e.g., \cite{Semmelmann:2002}:
\be\label{projection}
(\alpha_a)_\perp\equiv\alpha_a-\frac{1}{\pi+1}X_a\hook\bigl(e^b\wedge \alpha_b\bigr)-\frac{1}{n-\pi+1}e_a\wedge\bigl(X^b\hook\alpha_b\bigr)\,.
\ee
[Note that we automatically have $e^a\wedge (\alpha_a)_\perp=0=X^a\hook (\alpha_a)_\perp$ and hence $(\alpha_a{}_\perp)_\perp=(\alpha_a)_\perp$.] This projection naturally defines the {\em twistor operator} 
\be
K_a\omega\equiv (\nabla_a\omega)_\perp=\nabla_a\omega-\frac{1}{\pi+1}X_a\hook d\omega+ \frac{1}{n-\pi+1}e_a\wedge\delta \omega\,.
\ee
Using this notation we can rewrite Eq. \eq{ckk} as $\bigl(\nabla_a\omega+ \{B,\omega_a\}\bigr)_\perp=0$, or, 
\be\label{CKYeq}
K_a\omega+\{B,\omega_a\}_\perp=0 \,.
\ee  
This is the desired form of the {\em generalized conformal Killing--Yano system}. For general flux $B$, it represents a coupled system
of linear first order partial differential equations for homogeneous parts of inhomogeneous form $\omega$. Its solution is a necessary and sufficient condition for \eq{R} to be satisfied in the second and first derivative order.
It is shown in the next section that if and only if $B$ is a combination of a function, 1-form, and 3-form, the 
 generalized conformal Killing--Yano system \eq{CKYeq} decouples,  without any additional restrictions on $\omega$ and $B$, into a system of independent equations for homogeneous parts $\omega_p$ of the form $\omega$.

In order that $L$, given by $\Omega$, \eq{Omega}, really be a symmetry operator for $\mD$, the solutions of \eq{CKYeq} have to satisfy additional conditions following from \eq{eq2}. One finds the following integrability conditions for the twistor operator:
\be
\!\!\!\!\!\!\!\!\!\!\!\!\!\!\!\!\!\!\!\!\!\!\!\!\!\!\!\!
2e^b\wedge X^a\hook \nabla_{[a}K_{b]}\omega=e^b\wedge X^a\hook R(X_a,X_b)\omega-\frac{\pi}{\pi+1}\delta d\omega-\frac{n-\pi}{n-\pi+1}d \delta \omega\,.
\ee  
Taking further into account that 
$
2e^b\wedge X^a\hook \nabla_{[a}\{B,\omega_{b]}\}_\perp=\nabla^b \{B,\omega_{b}\}_\perp\,,
$
we arrive at the integrability conditions of Eq. \eq{CKYeq}
\be
\!\!\!\!\!\!\!\!\!\!
e^b\wedge X^a\hook R(X_a,X_b)\omega=\frac{\pi}{\pi+1}\delta d\omega+\frac{n-\pi}{n-\pi+1}d \delta \omega-\nabla^b \{B,\omega_{b}\}_\perp\,.
\ee
Using further the explicit form of $\Omega$, \eq{Omega}, in the term $e^a\nabla_a\Omega$, we can rewrite  Eq. \eq{eq2} as
\ba
\!\!\!\!\!\!\!\!\!\!\!\!\!\!
\delta\left(\frac{1}{n-\pi}X^b\hook \{B,\omega_b\}\right)-d\left(
\frac{1}{\pi}e^b\wedge\{B,\omega_b\}\right)\nonumber\\
\qquad
-\nabla^a\{B,\omega_a\}+\bigl\{B,\Omega\}-2(\eta\omega^a)\nabla_aB-\Sigma
B+d\f-\delta\ep=0\,. \label{integ}
\ea
It is now obvious, using the explicit form of $\Sigma$, \eq{sigma}, that this equation 
represents (for general $B$) an {\em additional first order partial differential constraint} on inhomogeneous form $\omega$;
we shall sometimes refer to the l.h.s. as {\em anomalies}.   
In principle, one could plug here the explicit expressions for
$\Omega$ and $\Sigma$, for general $B$ the final expression is, however, not very
illuminating. In certain special cases considered in the next section a considerable simplification of this equation occurs. 
Let us also mention that in certain special cases one may choose the (so far arbitrary) forms $\f$ and $\ep$ to simplify the anomalies, see, e.g., the example of black hole background in minimal gauged supergravity in Sec.~4.2.

\section{Examples}

\subsection{Conformal Killing vectors}

As a straightforward check of the results we have obtained, we
consider the case where $\omega$ is a one-form. Equation \eq{CKYeq}
reduces to the requirement that $\omega^\sharp$ be a confomal Killing
vector. Making use of the integrability conditions, \eq{integ} (with $\f=0=\ep$) reduces
to
\begin{equation}
\mathcal{L}_{\omega^\sharp} B = -\frac{\delta \omega}{n}(\pi-1)B.
\end{equation}
This condition can also be deduced by requiring that the action
\begin{equation}
S = \int(\overline{\psi} D \psi + \overline{\psi} B \psi)\, d\mathrm{vol},
\end{equation}
from which the modified Dirac equation follows, is invariant under the
conformal symmetry generated by $\omega^\sharp$.

\subsection{Generalized Killing--Yano equations with torsion}

Let us now consider the case when the Dirac spinor is coupled to a skew-symmetric torsion and a $U(1)$ field.
We take $B$ as in \eq{AT}, i.e., 
\be
B=iA-\frac{1}{4}T\,,
\ee
where $A$ is a 1-form and $T$ a 3-form. In \cite{HouriEtal:2010a} it
was argued that the factor of $-1/4$ is natural when considering a
connection with torsion $T$. 
Expanding the bracket $\{B,\omega_a\}$ by using formulas \eq{formulas} and applying \eq{omegaa} and \eq{pr1} we immediately find 
\ba
\!\!\!\!\!\!\!\!\{B,\omega_a\}=X_a\hook\bigl(-2iA\cwedge{1}\omega+\frac{1}{2}T\cwedge{1}\omega-\frac{1}{12}T\cwedge{3}\omega\bigr)+\frac{1}{2}(X_a\hook T)\cwedge{1}\omega\,,\nonumber\\
\!\!\!\!\!\!\!\!
X^a\hook \{B,\omega_a\}=\frac{1}{2}T\cwedge{2}\omega\,,\nonumber\\
\!\!\!\!\!\!\!\!
e^a\wedge\{B,\omega_a\}=\pi\bigl(-2iA\cwedge{1}{\omega}+\frac{1}{2}T\cwedge{1}\omega-\frac{1}{12}T\cwedge{3}\omega\bigr)-T\cwedge{1}\omega\,,
\ea
and 
\be
\!\!\!\!\!\!\!\!\!\!\!\!\!\!\!\!\!\!\!\!\!\!\!\!
\{B,\omega_a\}_\perp=\frac{1}{\pi+1}X_a\hook(T\cwedge{1}\omega)-\frac{1}{2}\frac{1}{n-\pi+1}e_a\wedge (T\cwedge{2}\omega)+\frac{1}{2}(X_a\hook T)\cwedge{1}\omega\,.
\ee
Obviously, the last bracket preserves the rank of each homogeneous part of $\omega$. Therefore, Eqs. \eq{CKYeq} split into a set of uncoupled homogeneous equations for $p$-form components of form $\omega$. Moreover, introducing the torsion covariant derivative \eq{covariantT} and the two associated operations \eq{def2} 
we can rewrite \eq{CKYeq} as 
\be\label{Ttwistor}
K_a^T\omega\equiv \nabla_a^T\omega-\frac{1}{\pi+1}X_a\hook d^T\omega+\frac{1}{n-\pi+1}e_a\wedge \delta^T \omega=0\,,
\ee
where $K_a^T$ is a twistor operator with torsion. Therefore each $p$-form component of $\omega$ has to satisfy the generalized conformal Kiling--Yano equation with torsion introduced in \cite{KubiznakEtal:2009b}. One also finds
\ba
\!\!\!\!\!\!\!\!\!\!\!\!\!\!\!\!\!\!\!\!\!\!\!\!\!\!\!\!\!\!\!\!\!\!\!\!
\Omega=\frac{\pi-1}{\pi}d\omega-\frac{n-\pi-1}{n-\pi}\delta^T\omega+2iA\cwedge{1}\omega
+\frac{2-\pi}{2\pi}T\cwedge{1}\omega
-\frac{1}{2}T\cwedge{2}\omega\!+\!\frac{1}{12}T\cwedge{3}\omega\!+\!\f\!+\!\ep\,,\nonumber\\
\!\!\!\!\!\!\!\!\!\!\!\!\!\!\!\!\!\!\!\!\!\!\!\!\!\!\!\!\!\!\!\!\!\!\!\!
\Sigma=-\frac{2\eta}{n-\pi}\delta^T\omega-2\eta\ep\,.
\ea
Note that when $\Sigma=0$, i.e., for $\delta^T\omega=0$ and $\ep=0$, the operator
$L$ graded commutes with the Dirac operator $\mD$. Thereafter we can rewrite $L$ as follows:
\be\label{realsymetry}
\!\!\!\!\!\!\!\!\!\!\!\!\!\!\!\!\!\!
L=2X^a\hook \omega \nabla_a+\frac{\pi-1}{\pi}d\omega+2iA\cwedge{1}\omega
+\frac{2-\pi}{2\pi}T\cwedge{1}\omega
-\frac{1}{2}T\cwedge{2}\omega+\frac{1}{12}T\cwedge{3}\omega+\f\,,
\ee
which (up to arbitrary 0-form $f$) is the symmetry operator derived in \cite{HouriEtal:2010a} [Eq. (4.18)] in the case when $A=0$.

In general, the additional constraint \eq{integ} reduces (after some lengthy calculation) to 
\be
2i(dA)\cwedge{1}\omega+A^{(cl)}+A^{(q)}-d\f+\delta\ep=0\,, \label{maxtoran}
\ee
where 
\ba
A^{(cl)}&=&\frac{1}{\pi-1}d(d^T\omega)-\frac{1}{2}dT\cwedge{1}\omega-\frac{1}{n-\pi+3}T\wedge\delta^T\omega\,,\\
A^{(q)}&=&\frac{1}{n-\pi-1}\delta(\delta^T\omega)-\frac{1}{6(\pi+3)}T\cwedge{3}d^T\omega+\frac{1}{12}dT\cwedge{3}\omega\,,
\ea
are the `quantum' and `classical' anomalies introduced in \cite{HouriEtal:2010a}.
Note that the $U(1)$ and torsion anomalies decouple (there are no
mixed terms including $A$ and $T$ together). In the case where
$\omega$ is a homogeneous form of rank $p$, the first three terms in
\eq{maxtoran} must vanish independently, being of rank $p$, $p+2$ and
$p-2$ respectively; one can use the freedom of $\f$ and $\ep$ to simplify some of these terms. 
In the absence of torsion we have the algebraic condition discussed in a more special case in \cite{AcikEtal:2009}.
In the absence of $U(1)$ field we recover the conditions discussed in
\cite{HouriEtal:2010a}. We stress however that this result is stronger
that in \cite{HouriEtal:2010a} as we have shown 
uniqueness---\emph{any} symmetry operator of the Dirac equation with torsion is determined
by a (possibly inhomogeneous) form $\omega$ which obeys the
generalized conformal Killing Yano equation in every rank, and for
which the anomaly vanishes. 

It can be verified that equations \eq{maxtoran} (with $\f=0=\ep$) are satisfied by the generalized closed conformal Killing--Yano tensors with torsion 
in Kerr--Sen black hole spacetimes in all dimensions  \cite{Sen:1992, CveticYoum:1996bps, Chow:2008}
in the case when the torsion is identified with the 3-form flux $H$ \cite{HouriEtal:2010b}. 
Another interesting example of the geometry where all the conditions can be satisfied is the most general black hole 
spacetime \cite{ChongEtal:2005b} of minimal gauged supergravity when the 
torsion is identified with the dual of Maxwell field $T=*F/\sqrt{3}$ \cite{KubiznakEtal:2009b}.
In this case $\omega$ is a generalized Killing--Yano with torsion 3-form which obeys \eq{maxtoran} if $\ep=0$ and the 0-form $f$ is chosen
to be $f=-\frac{1}{12}T\wedge\!\!\!\!_{{\!}_3}\,\omega$. Using further the fact that in this spacetime $T\wedge\!\!\!\!_{{}_1}\,\omega=0=T\wedge\!\!\!\!_{{\!}_2}\,\omega$, from \eq{realsymetry} we recover the symmetry operator
\be
L=2X^a\hook \omega \nabla_a+\frac{3}{4}d\omega+2iA\cwedge{1}\omega\,
\ee
 for the massive minimally coupled with torsion Dirac equation obtained by Wu \cite{Wu:2009b}.

\subsection{Dirac equation with potential}
As another important example, we consider the symmetry operator of the Dirac equation with scalar potential, $B=iV$. We find  
\be
\{V,\omega_a\}=2X_a\hook (V \omega_e)\,,\quad \{V,\omega_a\}_\perp=0\,,
\ee
where $\omega_e$ is the even $Z_2$-homogeneous part of $\omega$.
This means that every $p$-form part of $\omega$ obeys a conformal Killing--Yano equation, $K_a\omega=0$, and in addition we require
(setting $f=0=\ep$) 
\ba\label{coajak}
\!\!\!\!\!\!\!\!\!\!\!\!
-2e^a\nabla_a(V\omega_e)+2V\Omega_o-2(\eta\omega^a)\nabla_aV+\frac{2\eta}{n-\pi}\delta\omega V=0\,.
\ea
We consider two cases: i) $\omega$ is $Z_2$-odd, $\omega=\omega_o$, and ii) $\omega$ is $Z_2$-even, $\omega=\omega_e$.
In the first case Eq. \eq{coajak} reduces to 
\be\label{conV1}
(dV)^\sharp\hook \omega-\frac{V}{n-\pi}\delta \omega=0\,.
\ee
In particular, for $V=m=const$ we require $\delta \omega=0$. In the second case Eq. \eq{coajak} reduces to 
\be\label{conV2}
dV\wedge\omega +\frac{V}{\pi}d\omega=0\,,
\ee
when $V=m$ we require $d\omega=0$. 
So, we have re-derived the well known fact \cite{BennCharlton:1997} that the 
symmetry operators of the massive Dirac equation are given in terms of 
Killing--Yano tensors of odd rank or in terms of closed conformal Killing--Yano tensors of even rank. 

Let us also comment here on the constants of motion for {\em classical trajectories}. 
In the $U(1)$ case, it was demonstrated in \cite{AcikEtal:2009} that conformal Killing--Yano forms $\omega$ which give rise to symmetry operators of the minimally coupled Dirac equation, i.e., those obeying \eq{maxtoran}, $(dA)\wedge_{\!\!\! _1}\omega=0$, generate not only the constants of {\em geodesic motion}, $\dot u=\nabla_u u=0$, but also provide quadratic in velocity invariants for classical (charged particle) trajectories,  
$\dot u\equiv\nabla_u u=u\hook (dA)$.
Similarly, we now demonstrate that  
conformal Killing--Yano forms $\omega$ which give rise to symmetry operators of the Dirac equation with potential generate quadratic in velocity invariants for the corresponding particle trajectories. For this purpose we consider the Lagrangian ${\cal L}=u\cdot u -V^2$; the equations of motion are 
\be\label{poteq}
\dot u\equiv \nabla_u u=-V(dV)^\sharp\,.
\ee 
Let us first consider $\omega=\omega_o$ and define $w=u\hook \omega$. Then we find (denoting by $\xi=-\frac{1}{n-\pi}\delta\omega$)
\ba
\dot w&=&\dot u\hook \omega+u\hook \nabla_u\omega=-V(dV)^\sharp\hook \omega+u\hook (u^\flat\wedge \xi)
\nonumber\\
&=&-\frac{V^2}{n-\pi}\delta\omega+u\cdot u \,\xi-u^\flat\wedge(u\hook \xi)\nonumber\\
&=&(V^2+u^2)\xi-u^\flat\wedge (u\hook \xi)=-u^\flat\wedge (u\hook \xi)\,.
\ea
In the second equality we have used \eq{poteq} and the conformal Killing--Yano equation, in the third we applied the condition \eq{conV1}, and in the last we used the freedom to set the Hamiltonian ${\cal H}=u\cdot u +V^2=0$. 
From the antisymmentry of $\omega$ it is obvious now that $w\cdot \dot w=0$ and hence $c\equiv w\cdot w$ is a quadratic in velocity constant of motion for trajectories \eq{poteq}. This constant corresponds to the rank-2 {\em conformal Killing tensor}  $K_{ab}=\omega_{ac}\omega^c_{\ b}$.
Similarly, when $\omega=\omega_e$ and the condition \eq{conV2} is satisfied one can show that $\tilde c\equiv (u^\flat\wedge \omega)\cdot (u^\flat\wedge \omega)$ is a quadratic in velocity constant of motion for trajectories \eq{poteq}.\footnote{We remark that when $V=const$ the corresponding conformal Killing tensor is of the {\em gradient type}, i.e., obeying $\tilde K_{(ab;c)}=g_{(ab}\tilde K_{c)}$ with $\tilde K_a=\nabla_a\tilde K$.} Therefore we have established that in both cases $\omega$ gives rise to constants of motion of classical trajectories
\eq{poteq}. 
An alternative way to see this would be to use the geometric optics approximation. 

\subsection{5-form flux}

For a $5$-form flux, the bracket $\{B, \omega_a\}_\perp$ has $(p+2)$-
and a $(p-2)$-form components:
\be
\{B, \omega_a\}_\perp = 2(B\cwedge{1}\omega_a)_\perp - \frac{1}{3}(B
\cwedge{3} \omega_a)_\perp\, .
\ee
Thus the generalized Killing--Yano equation \eq{CKYeq} does not split
into conditions on each rank of the inhomogeneous $\omega$, but rather
mixes the ranks. One may make an ansatz that $\omega$ is a homogeneous
form of rank $p$, in which case one must impose that $\{B,
\omega_a\}_\perp=0$ for consistency. The interpretation is then that a
standard conformal Killing--Yano tensor gives rise to a symmetry in the
case of a five-form flux, but that $\{B,\omega_a\}_\perp=0$ and
\eq{integ} are additional equations which must be satisfied in order
that this is the case.

\subsection{7-form flux}

Unlike in the case of the $5$-form flux, for $B$ a $7$-form one finds
that the bracket has $(p+4)$-, $p$- and $(p-4)$-form components:
\be
\{B, \omega_a\}_\perp = 2(B\cwedge{1}\omega_a)_\perp - \frac{1}{3}(B
\cwedge{3} \omega_a)_\perp+ \frac{1}{60}(B
\cwedge{5} \omega_a)_\perp\, .
\ee
Again, there will in general be mixing between different rank
components of the inhomogeneous $\omega$. If we seek a homogeneous
$\omega$, then \eq{CKYeq} consists of the two algebraic conditions
\be
(B\cwedge{1}\omega_a)_\perp=0, \qquad (B\cwedge{5} \omega_a)_\perp=0\, ,
\ee
together with a modified Killing--Yano equation which may be formally written as
\be\label{twistor2}
K^B_a \omega = 0\,,
\ee
where $K^B_a$ is the twistor operator of the `connection'
\be
\nabla^B_a \omega = \nabla_a \omega + \frac{1}{3}(X_a \hook B)
\cwedge{3} \omega\, 
\ee
and $\delta^B=-X^a\hook \nabla_a^B$ and $d^B=e^a\wedge \nabla_a^B$ are the associated two operations.
Note, however, that $\nabla^B$ does not respect the wedge product or Clifford product and so it is not
a standard connection on the exterior bundle.
On the other hand, one has $*\nabla^B=\nabla^B *$ and hence the modified twistor equation \eq{twistor2} is invariant under the Hodge duality;
if $\omega$ solves Eq. \eq{twistor2} so does $*\omega$. 

. 

\section{Discussion and conclusions}

In this paper we have considered the problem of finding symmetry
operators for a Dirac equation coupled to arbitrary $p$-form
fluxes. As a result, we have been able to characterize all first order
symmetry operators; they are given in terms of an inhomogeneous form $\omega$ subject to various algebraic and differential constraints,
Eqs. \eq{CKYeq} and \eq{integ}.
 The main application which we have presented is to
show that the operators previously constructed for the Dirac operator
with a $3$-form torsion in
\cite{HouriEtal:2010a} are essentially unique. We were further able to
include a minimal coupling to a $U(1)$ Maxwell field into this
analysis. 
As a completely new application we have considered the Dirac equation in scalar potential and Dirac equation in the presence of 
$5$- and $7$-form fluxes.

In the case of a 7-form flux we were able to define a new `connection' and the corresponding generalization of the twistor equation. Such an equation has to be satisfied by a form $\omega$ determining the symmetry operator in the case when this form is {\em homogeneous}. This case has a direct generalization for fluxes of the rank $p=4k+3$, where $k=0,1,2,\dots$ One can easily show that homogeneous form $\omega$ has to satisfy the {\em modified twistor equation}
\be\label{twistor3}
K^B_a \omega = 0\,,\quad \nabla_a^B\omega=\nabla_a\omega-2\frac{(-1)^k}{(2k+1)!}(X^a\hook B)\cwedge{2k+1}\omega\,.
\ee   
(We stress, however, that the new `covariant derivative' on forms $\nabla^B_a$ does not obey the Leibnitz rule with respect to the wedge product or Clifford product, unless $k=0$.) Similar to \eq{twistor2}, the modified twistor equation \eq{twistor3} is {\em invariant} under the Hodge duality.
For fluxes of the rank $p\neq 4k+3$ the homogeneous form $\omega$ has to satisfy the standard twistor equation $K_a\omega=0$. In both cases the (modified) twistor equation is accompanied with additional algebraic conditions following from the requirement that all terms of $\{B,\omega_a\}_\perp$ which are not of the same rank as $\omega$ have to vanish as well as with an additional first-order differential constraint given by Eq. \eq{integ}.
In the case when $\omega$ is allowed to be {\em inhomogeneous}, the required equations \eq{CKYeq} and \eq{integ} represent a coupled system of equations. In consequence, the restrictions on each homogeneous part of $\omega$ are much weaker.  

Let us finally emphasize that although the requirements \eq{CKYeq} and \eq{integ} seem very restrictive there are non-trivial examples of supergravity backgrounds where these are satisfied. This is for example the case of spacetimes with $U(1)$ and torsion fluxes---such as Kerr--Sen  geometries in all dimensions or the most general spherical black hole spacetime of minimal gauged supergravity.  It is an interesting open question whether one can find analogous symmetries in backgrounds with fluxes of higher-rank, 5-form flux for example. 

\section*{Acknowledgments}

We would like to thank Y.~Yasui for helpful comments and M.~Cariglia and T. Houri for reading the manuscript. D.K. acknowledges the Herschel Smith Postdoctoral Research Fellowship
at the University of Cambridge. P.K. thanks 
DAMTP for the kind hospitality during his stay in Cambridge and 
the grant GA\v{C}R 202/08/0187.

\appendix

\section{Notations}
In this appendix we gather our conventions and formalism, these are essentially taken from \cite{BennTucker:book, HouriEtal:2010a}.
$M$ is a $n$-dimensional (pseudo)-Riemannian manifold equipped with a metric $g$, 
$\{{X}_a \}$  denotes an orthonormal basis for $TM$,
${g}({X}_a,{X}_b)=\eta_{ab}$, and $\{{e}^a\}$ is a dual basis
for $T^*M$ with ${g}({e}^a,
{e}^b)=\eta^{ab}$. We additionally define
\be
{X}^a = \eta^{ab}{X}_b\,, \qquad {e_a} = \eta_{ab} {e}^b.
\ee
Operations $\flat$ and $\sharp$ correspond to `lowering' and `rising' of indices of vectors and  forms, respectively.
We shall further make use of
the $n$-fold contracted wedge product defined for any $p$-form ${\alpha}$ and $q$-form ${\beta}$ inductively by \cite{HouriEtal:2010a}
\be
{\alpha}
\cwedge{0} {\beta} ={\alpha}
\wedge {\beta}\,, \qquad {\alpha}
\cwedge{k} {\beta} = ({X}^a \hook {\alpha}) \cwedge{k-1} ({X}_a
\hook {\beta})\,,
\ee
where the `hook' operator $\hook$ corresponds to the inner derivative.
For an arbitrary $p$-form $\alpha$ this product satisfies
\be\label{pr1}
X_a \hook (\alpha \cwedge{k}\beta) = (-1)^{k} (X_a \hook \alpha)
\cwedge{k} \beta + (-1)^p \alpha \cwedge{k} (X_a \hook \beta)\,.
\ee
Throughout the paper $\nabla$ denotes the Levi-Civita connection and we use a shorthand $\nabla_a\equiv \nabla_{X_a}$. The exterior derivative and co-derivative on forms are  
\be\label{dd}
d=e^a\wedge \nabla_a\,,\quad \delta=-X^a\hook \nabla_a\,.
\ee
With respect to the contracted wedge product, $\nabla_a$ is a derivation
\be\label{A22}
\nabla_a(\alpha \cwedge{k}\beta) = \nabla_a\alpha \cwedge{k}\beta+
\alpha \cwedge{k}\nabla_a\beta\,,
\ee
and, when $\alpha \in \Lambda^p(M)$, then
\begin{eqnarray}
\delta(\alpha \cwedge{k}\beta) &=& (-1)^k \delta \alpha \cwedge{k}
\beta - (-1)^p \nabla_a \alpha \cwedge{k} (X^a \hook \beta) \nonumber \\
&& \quad +(-1)^p  \alpha \cwedge{k}
\delta \beta - (-1)^k (X^a \hook \alpha) \cwedge{k} \nabla_a \beta\,,\label{A23}\\
d(\alpha \cwedge{k}\beta) &=& (-1)^k d \alpha \cwedge{k}
\beta - (-1)^k k \nabla_a \alpha \cwedge{k-1} (X^a \hook \beta) \nonumber \\
&& \quad +(-1)^p\alpha \cwedge{k}
d\beta - (-1)^p k (X^a \hook \alpha) \cwedge{k-1} \nabla_a \beta \,.\label{A24}
\end{eqnarray}

When working with spinors, we identify the elements of the Clifford algebra with differential forms and denote the Clifford multiplication by juxtaposition. This is defined for a 1-form $\alpha$ and $p$-form $\omega$ by 
\be\label{Cliff}
\alpha\omega=\alpha\wedge \omega +\alpha\cwedge{1}\omega\,,\quad
\omega\alpha=(-1)^p\bigl(\alpha\wedge \omega -\alpha\cwedge{1}\omega\bigr)\,.
\ee  
By repeating the application of this rule we construct the Clifford product between forms of arbitrary degree. 
Let $\alpha \in \Lambda^p(M)$ and $\beta \in \Lambda^q(M)$.
Then, 
\be\label{formulas}
\!\!\!\!\!\!\!\!\!\!\!\!\!\!\!\!\!\!\!\!\!\!\!\!\!\!\!\!\!\!\!\!\!\!\!\!
\alpha \beta=\sum_{k=0}^{p}\frac{(-1)^{k(p-k)+[k/2]}}{k!}\alpha\cwedge{k}\beta\,,\quad
\beta\alpha=(-1)^{pq}\sum_{k=0}^{p}\frac{(-1)^{k(p-k+1)+[k/2]}}{k!}\alpha\cwedge{k}\beta\,,
\ee
with $[k/2]$ being the integer part of $k/2$.
We further define the following {\em bracket} used in the paper, Eq. \eq{bracketpaper}:
\be\label{bracket}
\{\alpha,\beta\}\equiv\alpha\beta-(-1)^{q}\beta\alpha\,.
\ee
(Note that unless $p$ is odd, this bracket differs from the {\em Clifford graded commutator},
$[\alpha,\beta]\equiv\alpha\beta-(-1)^{pq}\beta\alpha$, used in \cite{AcikEtal:2009}.)
In terms of more familiar $\gamma$-matrices, the identification between
differential forms and elements of the Clifford algebra is expressed as
\begin{equation}
\omega=\frac{1}{p!}\omega_{a_1 \ldots a_p} e^{a_1} \wedge \cdots \wedge
e^{a_p} \longrightarrow \frac{1}{p!}\omega_{a_1 \ldots a_p}
\gamma^{[a_1} \cdots \gamma^{a_p]}. \label{gammacor}
\end{equation}
Evidently, the relations (\ref{Cliff}) are equivalent to the usual
anti-commutator for the $\gamma$-matrices
$\gamma^a \gamma^b + \gamma^b \gamma^a = 2 \eta^{ab}$.

We shall also work with inhomogeneous forms $\alpha$ which we decompose either into their odd, $\alpha_o$, and even, $\alpha_e$, 
$Z_2$-homogeneous parts, $\alpha=\alpha_o+\alpha_e$, or into their $p$-form homogeneous parts $\alpha_p$ as follows:
\be
\alpha=\sum_{p=0}^n \alpha_p\,,\quad \alpha_p\in \Lambda^p(M)\,.
\ee
For such forms we define the following two operations:
\be
\pi\alpha\equiv\sum_{p=0}^n p\alpha_p\,,\quad \eta\alpha\equiv\sum_{p=0}^n (-1)^p\alpha_p\,.
\ee
It is obvious that all the above operations naturally extend to inhomogeneous forms. In particular, for the bracket \eq{bracket} we have 
\be\label{bracket2}
\{\alpha,\beta\}=\alpha\beta-(\eta\beta)\alpha\,.
\ee
One also has
$e^a\wedge (X_a\hook \alpha)=\pi\alpha$ and $X_a\hook (e^a\wedge \alpha)=(n-\pi)\alpha$.

Finally, we shall need the curvature operator
\be
R(X_a,X_b)=2\nabla_{[a}\nabla_{b]}-\nabla_{[X_a,X_b]_{LB}}\,,
\ee
where $[\ ,\ ]_{LB}$ denotes the Lie bracket of two vector fields.
When acting on a spinor, it is given by \cite{BennTucker:book}
\be\label{Rspinor}
R(X_a,X_b)\psi={\cal R}_{X_a,X_b}\psi\,,
\ee
where ${\cal R}_{X_a,X_b}$ is a 2-form associated with the usual curvature 2-form $R_{ab}$ of the Levi-Civita connection $\nabla$, 
${\cal R}_{X_a,X_b}=-\frac{1}{4}X_a\hook X_b\hook R_{cd}e^{cd}$. To simplify our calculations we shall freely use a basis which is parallel at a point, i.e., 
which satisfies 
\be
\nabla_a X_b=\nabla_ae^b=[X_a,X_b]_{LB}=0\,. 
\ee

\hspace{3cm}

\providecommand{\href}[2]{#2}\begingroup\raggedright\endgroup


\end{document}